\documentclass{article}

\usepackage{arxiv}

\usepackage[utf8]{inputenc} % allow utf-8 input
\usepackage[T1]{fontenc}    % use 8-bit T1 fonts
\usepackage{hyperref}       % hyperlinks
\usepackage{url}            % simple URL typesetting
\usepackage{booktabs}       % professional-quality tables
\usepackage{amsfonts}       % blackboard math symbols
\usepackage{nicefrac}       % compact symbols for 1/2, etc.
\usepackage{microtype}      % microtypography
\usepackage{lipsum}		% Can be removed after putting your text content
\usepackage{graphicx}
\usepackage{doi}
\usepackage{bbm}
\usepackage{amssymb}
\usepackage{amsmath}
\usepackage{hyperref}
\usepackage{comment}

\usepackage{algorithm}
\usepackage{algorithmic}

\usepackage{lineno}
\title{The Topology and Geometry\\of Neural Representations}

\author{Baihan Lin\\
	Department of Neuroscience\\
	Columbia University\\
	New York, NY 10027 \\
	\texttt{bl2681@columbia.edu} \\
	\And
	Nikolaus Kriegeskorte \\
	Department of Neuroscience\\
	Columbia University\\
	New York, NY 10027 \\
	\texttt{nk2765@columbia.edu} \\
}

\renewcommand{\headeright}{}
\renewcommand{\undertitle}{}
\renewcommand{\shorttitle}{The Topology and Geometry of Neural Representations}

\hypersetup{
pdftitle={The Topology and Geometry of Neural Representations},
pdfsubject={q-bio.NC},
pdfauthor={Baihan Lin, Nikolaus Kriegeskorte},
pdfkeywords={Representational similarity analysis, Representational topology, Representational geometry},
}

\begin{document}
\maketitle

\begin{abstract}
A central question for neuroscience is how to characterize brain representations of perceptual and cognitive content. An ideal characterization should distinguish different functional regions with robustness to noise and idiosyncrasies of individual brains that do not correspond to computational differences. Previous studies have characterized brain representations by their representational geometry, which is defined by the representational dissimilarity matrix (RDM), a summary statistic that abstracts from the roles of individual neurons (or responses channels) and characterizes the discriminability of stimuli. Here we explore a further step of abstraction: from the geometry to the topology of brain representations. We propose topological representational similarity analysis (tRSA), an extension of representational similarity analysis (RSA) that uses a family of geo-topological summary statistics that generalizes the RDM to characterize the topology while de-emphasizing the geometry. We evaluate this new family of statistics in terms of the sensitivity and specificity for model selection using both simulations and functional MRI (fMRI) data. In the simulations, the ground truth is a data-generating layer representation in a neural network model and the models are the same and other layers in different model instances (trained from different random seeds). In fMRI, the ground truth is a visual area and the models are the same and other areas measured in different subjects. Results show that topology-sensitive characterizations of population codes are robust to noise and interindividual variability and maintain excellent sensitivity to the unique representational signatures of different neural network layers and brain regions. These methods enable researchers to calibrate comparisons among representations in brains and models to be sensitive to the geometry, the topology, or a combination of both.
\end{abstract}

% keywords can be removed

\section{Introduction} 

\begin{figure}[tb]
% \vspace{-0.8cm}
\centering 
    \includegraphics[width=\linewidth]{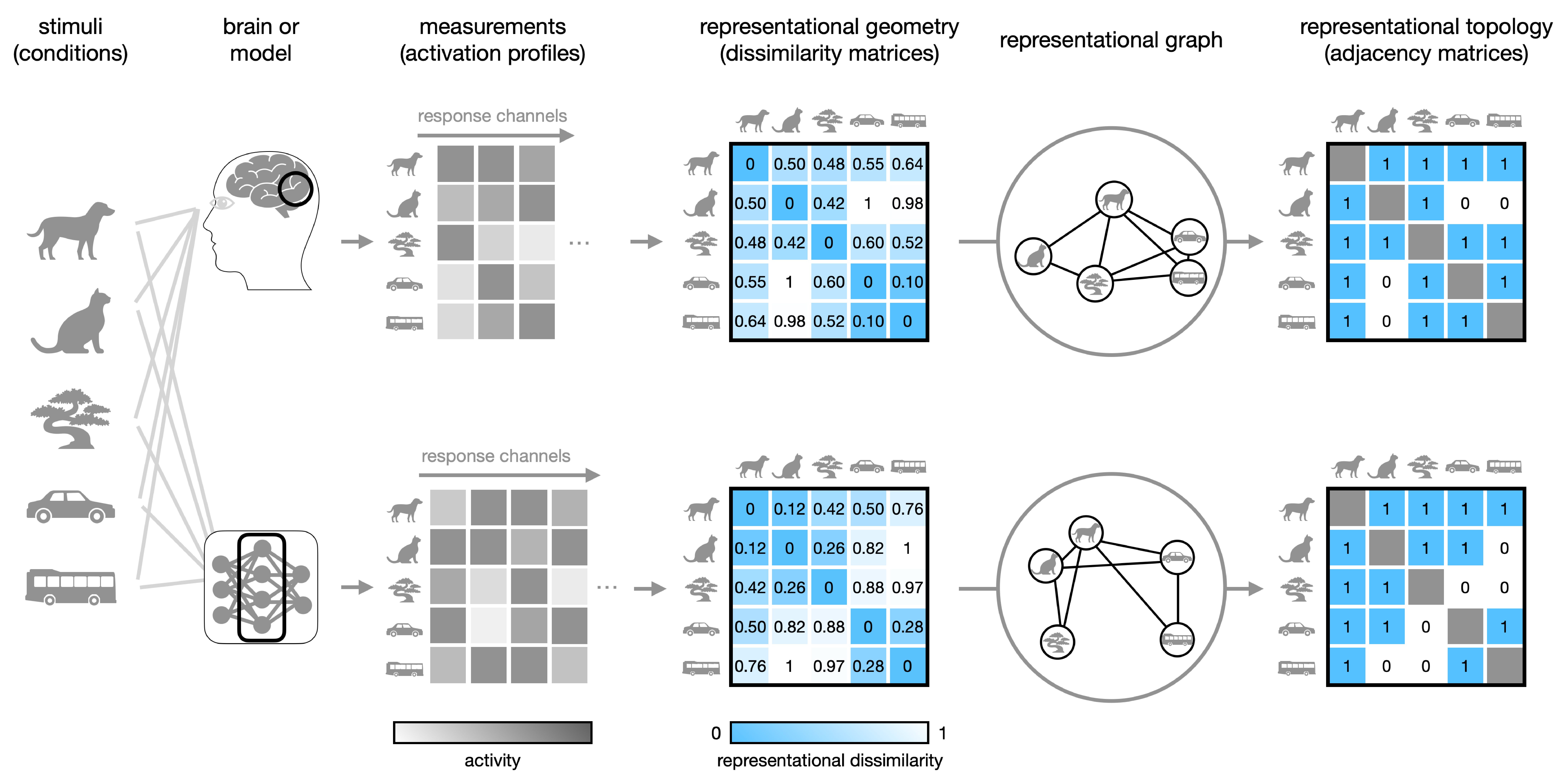}
\caption{\footnotesize \textbf{Comparing representations between brains and models.} To understand the degree to which a computational model can account for the cognitive process of a certain brain region, the same set of stimuli is presented to both the model and the biological system. The response patterns across measured response channels (e.g. neurons or voxels) are then characterized by a summary statistic, the representational dissimilarity matrix (RDM, center), which defines the metric configuration of the stimuli in the neural population response space. However, the metric configuration can be sensitive to measurement noise and idiosyncrasies of individual brains that do not reflect computational function. An alternative summary statistic that captures the topology would be the adjacency matrix (right), which defines the unweighted graph of neighborhood relationships in the population response space. This summary statistic promises to be more robust to noise and idiosyncrasies, but may discard too much information. Considering the geometry (RDM) and topology (adjacency matrix) as extremes of a continuum suggests that it may be possible to get the best of both (Fig. 2).
}\label{fig1}
\end{figure}

Geometrical and topological analyses can be applied profitably to the structure and connectivity of brains, on the one hand, and to neural population code representations on the other. Let us first consider the realm of brain structure and connectivity. Geometrical characterizations have been used to study the physical structure of brains (e.g. the geometry of the cortical surface \cite{dale1999cortical,desikan2006automated,fischl2002whole,kriegeskorte2001efficient}) and anatomical connectivity \cite{bastos2016tutorial,honey2009predicting,van2010exploring}. Topological and graph-based characterizations have been extensively used in network neuroscience \cite{bassett2017network,medaglia2017brain,gu2015controllability,srivastava2020models,bullmore2012economy,farahani2019application,sporns2022graph} to investigate the anatomical connections and functional correlations between brain regions and how structural and functional network topology is related to cognition. This paper is not about either the physical structure or the connectivity of the brain, but about the topology and geometry of neural representations.

In the realm of the neural representations, geometrical characterizations have been used to study the relationships between neural population activity patterns: the representational geometry \cite{kriegeskorte2008matching,kriegeskorte2008representational,kriegeskorte2013representational,kriegeskorte2021neural,sorscher2022neural,nieh2021geometry,freeman2018neural,chung2021neural} (Fig. \ref{fig1}, left, middle). The representational geometry provides a useful intermediate level of description capturing the information represented in a neuronal population code, while abstracting from the roles of individual measured responses (reflecting neurons or voxels) \cite{kriegeskorte2019peeling}. Considering the representational dissimilarities, rather than the representational patterns, enables direct comparisons of population-code representations between different individuals and species, as well as between brains and computational models. The analysis of representational geometries, known as representational similarity analysis (RSA, \cite{kriegeskorte2008representational}), has been successfully applied to understand diverse functions \cite{kriegeskorte2013representational}, including perception in the visual \cite{kriegeskorte2008matching}, auditory \cite{sievers2021visual} and other modalities \cite{fournel2016multidimensional} and higher cognitive functions such as abstraction \cite{nieh2021geometry,chung2021neural}, decision-making \cite{van2019computational}, working memory, social cognition \cite{thornton2018theories, thornton2023brain, tamir2016neural, tamir2018modeling}, and planning \cite{ehrlich2022geometry}. Representational geometries can be visualized by arranging stimuli in two or three dimensions, such that their distances approximately reflect the corresponding distances in the high-dimensional neural response space. Representational geometries, captured by the matrix of pairwise distances (the representational dissimilarity matrix, RDM), can also be used as a basis for model comparison \cite{kriegeskorte2008representational,nili2014toolbox,schutt2023statistical}, an approach that has enabled researchers to adjudicate among competing models of brain representations \cite{khaligh2014deep,kietzmann2019recurrence,cichy2017multivariate,konkle2022can}.

When we investigate the representational geometry by considering distances among neural activity patterns, we abstract from the roles of individual neurons. The representational topology provides a further step of abstraction. We may care less about the precise distances among the points in the high-dimensional response space that define the geometry than about the way the points hang together in what is sometimes called the neural manifold \cite{low2018probing,kriegeskorte2021neural,chung2021neural}. We may hypothesize, for example, that the overall geometry of the representation in a given cortical area or layer of a neural network model may vary across individual people or instances of a neural network model trained from different random seeds \cite{mehrer2020individual}. If the corresponding cortical areas in two people or the corresponding layers in two model instances served the same computational purpose, however, we may expect that stimuli that are neighbors in one individual's (or model instance's) representation remain neighbors in the other individual's (or instance's) representation. A graph of representational neighborhood relationships can be obtained by thresholding the distance matrix (Fig. \ref{fig1}, right). The thresholding operation is well-motivated when we care only about whether two points are in the same neighborhood or not. If they are in the same neighborhood, we consider them related and do not care whether they are close or very close. If they are not in the same neighborhood, we consider them unrelated and do not care whether they are very far apart or merely far enough not to count as neighbors. The neighborhood graph characterizes the representational topology.

An important question is to what extent the further step of abstraction involved in going from the distance matrix (geometry) to the neighbor graph (topology) is desirable or undesirable. It could be desirable for providing a more robust reduced signature of a region's computational function. However, it could be undesirable if it removes geometrical information important for discerning regions that implement distinct computational functions. Here we address this question empirically, using human functional MRI data and simulations based on neural network models.

Topological data analysis techniques \cite{wasserman2018topological} such as the persistent homology \cite{edelsbrunner2008persistent} and the Mapper algorithm \cite{singh2007topological}  are popular in many fields of biology \cite{rizvi2017single,ttda,geniesse2019generating,bibm,ellis2019feasibility,pike2020topological} and have also been used to directly analyze the representational space of the population activity \cite{chaudhuri2019intrinsic}. For instance, a study has discovered that the structure of spontaneous and evoked activity patterns in V1 can be mapped onto a manifold that has the topology of a sphere, whose two dimensions may reflect orientation and spatial frequency \cite{singh2008topological}, with the population response selective to the extremes of spatial frequency mapped towards the two poles of the sphere. Similarly, a recent study applied topological analysis techniques to study the population activity of grid cells, which are thought to be involved in spatial navigation and orientation \cite{gardner2022toroidal}. This study found that the population activity of grid cells has a toroidal topology, meaning that the manifold wraps around like a donut, whose two surface dimensions correspond to the 2d space navigated, implementing a cyclic representation. These examples illustrate the power of topological data analysis techniques to reveal the structure of the neighborhood graph of neural population representations.

These inspiring studies notwithstanding, topological characterizations are more widely used in network neuroscience and only beginning to impact investigations of the relationships among neural population activity patterns. Here we build on the early topological analyses of neural population activity patterns \cite{singh2008topological,ellis2019feasibility,gardner2022toroidal,low2018probing} and introduce a new family of summary statistics that can characterize the geometry as well as the topology of neural activity patterns. These geo-topological summary statistics enable researchers to calibrate the geometric and topological sensitivity of the analysis, so as to define a good signature of the computational role of each brain region. The new representational signatures can then be used not only for visualization of the representational geometry and topology, but also as a basis for formal inferential model comparison in the framework of RSA \cite{schutt2023statistical}, where our geo-topological summary statistics can replace the RDM, which characterizes the geometry.

Consider the example of visual perception (Fig. \ref{fig1}). We begin by measuring the brain-activity patterns elicited by each of a set of stimuli in a brain region or computational model. By estimating the distances among the stimulus representations (with full metric information), we can gain insights into distinctions the brain region or model layer emphasizes. Metric distance estimates promise detailed geometrical information, but are sensitive to noise and individual idiosyncrasies. Thresholding the distances provides a graph with binary edges, which captures how the neural manifold hangs together and also promises to be more resilient to nuisance variation. The methods we introduce here share a focus on neighborhood relationships with popular visualization techniques like Isomap \cite{tenenbaum2000global}, locally linear embedding \cite{roweis2000nonlinear}, $t$-SNE \cite{vanDerMaaten2008tSNE}, and UMAP \cite{mcinnes2018umap}. However, while these techniques aim to visualize a single representation, our aims are to characterize multiple representations in models and brains, quantify their similarity, and statistically compare models in terms of their ability to account for the topology and geometry of brain representations. To combine the benefits of detailed metric information and a binary edge description, we seek to define a representational graph that captures aspects of both the representational geometry and topology using weighted edges.

To integrate geometric (distance-based) and topological (graph-based) characteristics, we define the edge weights of the graph by a nonlinear monotonic transformation of the distances that (1) emphasizes the distinction between small and large distances, (2) compresses very small distances, thus disregarding the distinctions among them, (3) represents a continuum of intermediate distances, and (4) compresses very large distances. An example of the effectiveness of this transformation is an adaptive generalization of distance correlation based on proposed transformed distances, which provides a dependence measure with robust sensitivity to geometric and topological characteristics \cite{lin2018adaptive,www2022}. 

There are two motivations for defining the edge weights as a nonlinear monotonic transform of the distances, one theoretical and one data-analytical. From a theoretical perspective, differences among very large representational distances may not provide the most useful signature of the computational function of a brain region. It is the local geometry that determines which stimuli the representation renders indiscriminable, which it discriminates, but places together in a cluster, and which it places in different neighborhoods. The global geometry of the clusters (whether two stimuli are far or very far from each other in the representational space) may be less relevant to computation for two reasons. First, once two stimuli are perfectly discriminable, moving them even further apart does not improve discriminability. Second, in a high-dimensional space, a set of randomly placed clusters will tend to afford linear separability of arbitrary dichotomies among the clusters \cite{rigotti2013importance,kushnir2018neural,cover1965geometrical} independent of the exact global geometry. 

Like in a storage room, related things may need to be placed together in a representational space and unrelated things in different locations. The requirement of co-localization strongly constrains the local geometry because there is only one direction toward a given location. The requirement that two things be far from each other, by contrast, only weakly constrains the global geometry, because there are many directions away from a given location, especially in a high-dimensional space. This argument suggests the hypothesis that variations among large distances are idiosyncratic to an individual brain or model instance \cite{mehrer2020individual}. If this hypothesis were true, then compressing this variation may help us focus on more functionally relevant features of the representation that are less variable across individuals and model instances.

\begin{figure}[tb]
% \vspace{-0.8cm}
\centering
    \includegraphics[width=.8\linewidth]{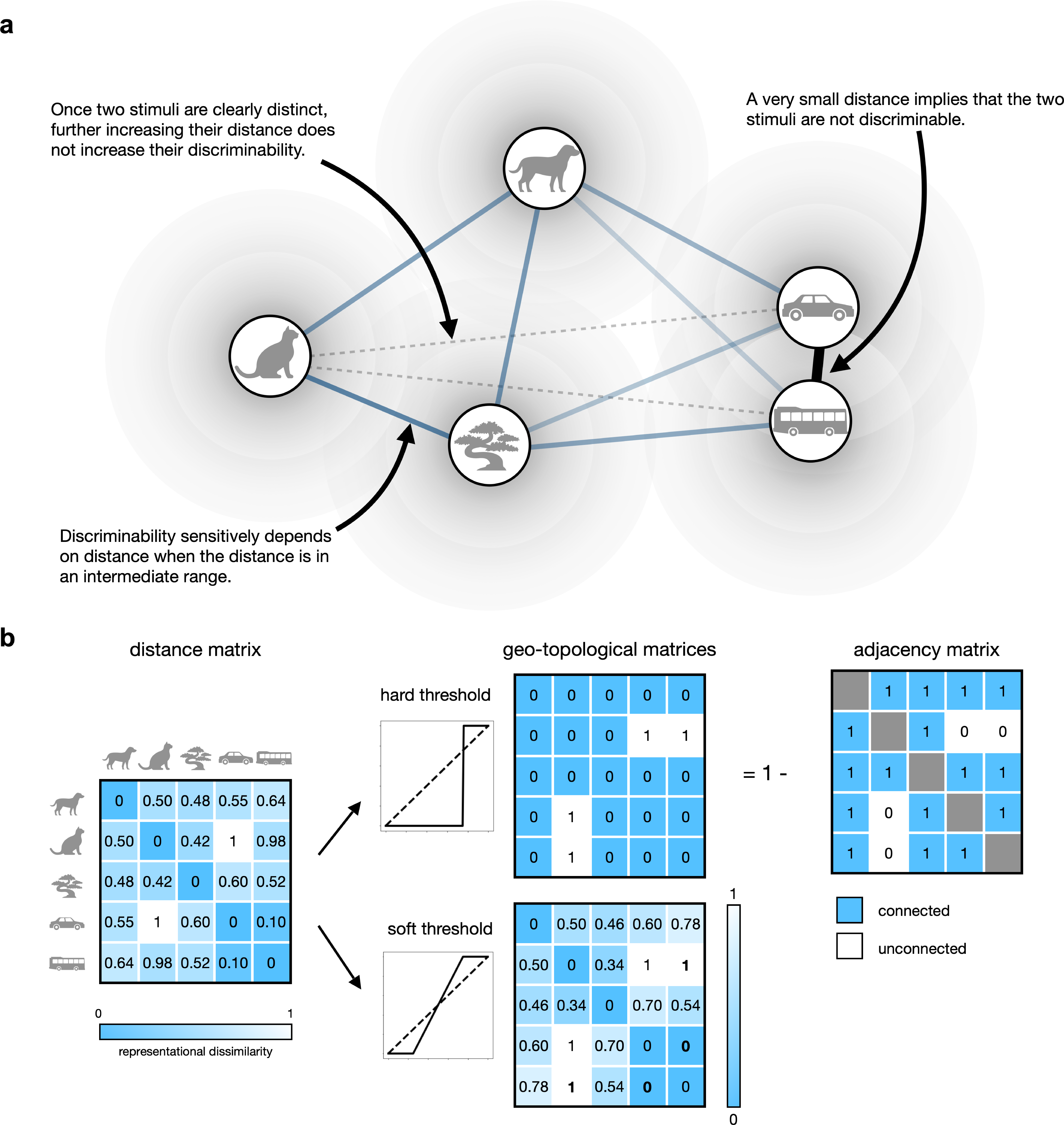}
\caption{\footnotesize \textbf{Intuition of the geo-topological transform of distances.} \textbf{(a)} Consider the five visual stimuli from Fig. \ref{fig1}, whose representation in a visual cortical area can be characterized by its representational geometry. If the response patterns are all affected by isotropic noise (or the noise has been whitened by a transform), then the Euclidean distances monotonically reflect the discriminabilities. Variation among large distances, however, is not associated with great differences in discriminability, because all pairs of well-separated stimuli are nearly perfectly discriminable (dashed lines). Similarly, variation among very small distances is not associated with great differences in discriminability, because all pairs of neighboring stimuli are indiscriminable (thick black edge). This suggests that variation among small distances and variation among large distances can be suppressed in favor of emphasizing the transition from small to large distances.  \textbf{(b)} To emphasize the transition from small to large distances while suppressing variation among small distances and variation among large distances, we can threshold the distances, such that small distances are pushed to zero and large distances are pushed to the maximum. We can either use a hard threshold (upper row) or a soft threshold (lower row). A hard threshold yields a binary matrix whose complement is the adjacency matrix of a graph that connects neighboring stimuli. A soft threshold creates a continuous transition, reflecting the graded increase in discriminability as the distance grows, and defines a weighted graph, where the weights reflect distances, but the pairs of stimuli that are furthest from each other are not directly connected (dashed lines in (a)). 
}\label{fig2}
\end{figure}

From a data-analytical perspective, conversely, very small distances may be unreliable given the various noise sources that affect the measurements. Compressing small distances, thus, promises to reduce the influence of measurement noise on visualizations and inferential results. 
Compressing small and large distances is achieved by thresholding of the distances. However, thresholding may be too aggressive in that it completely removes all continuous information reflecting the geometry. As illustrated in Fig. \ref{fig2}, on one end, we have the distance matrix with the full metric information, and on the other, we simply have an adjacency matrix telling us whether two stimuli are neighbors in the representational space or not. To get the best of both worlds, it seems attractive to focus our sensitivity on a particular intermediate range of distances, so as to maintain reliable geometric information, while reducing the influence of noise (by compressing variation among small distances) and the influence of individual idiosyncrasies (by compressing variation among large distances). We show that this can be accomplished by a monotonic transform of the representational distances and that the resulting geo-topological representational summary statistics robustly reveal the functional distinctions among human brain regions and DNN layers. We introduce a family of geo-topological summary statistics that generalizes the RDM and provides a basis for topological RSA (tRSA), a generalization of RSA that balances sensitivity to the topology and geometry of neural representations.

\section{Materials and Methods}

\begin{figure}[tb]
% \vspace{-0.8cm}
\centering
    \includegraphics[width=.8\linewidth]{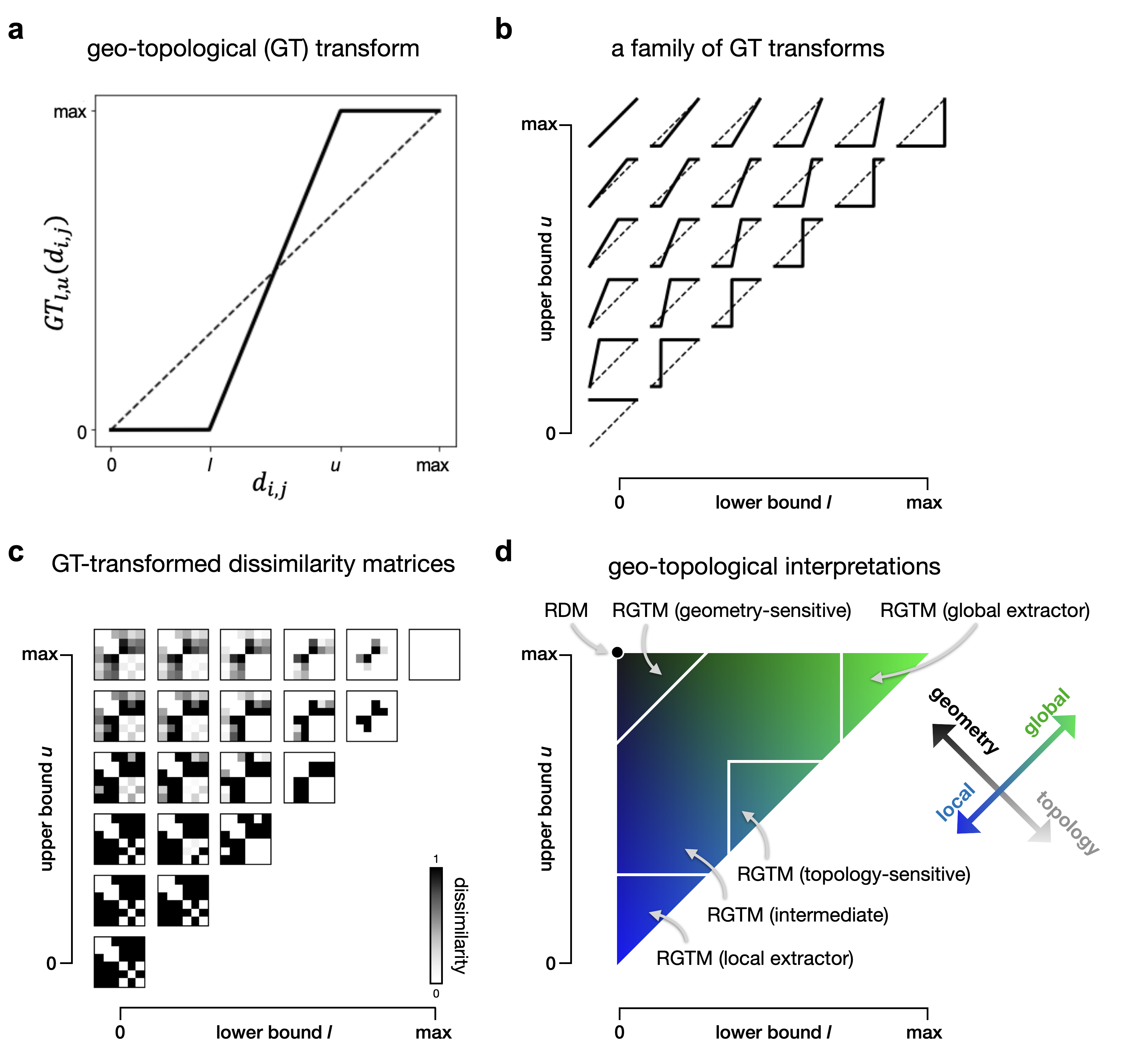}
\caption{\footnotesize \textbf{A family of geo-topological transforms of the RDM.} \textbf{(a)} The geo-topological (GT) transform is formulated as a linear piecewise function, such that any distances smaller than a lower bound $l$ will be mapped to zero, and any distances bigger than an upper bound $u$ will be mapped to 1. Between $l$ and $u$, the transition is linear. \textbf{(b)} By varying the thresholds $l$ and $u$, we select among a family of GT transforms. \textbf{(c)} By applying different GT transforms to the RDM, we obtain so-called representational geo-topological matrices (RGTMs). \textbf{(d)} To interpret the way the GT transforms reflect geometric and topological properties of the representation, we group the family members in different zones of the plane spanned by $l$ and $u$. The closer a GT transform is to the upper left corner ($l=0, u=max$), the more similar the RGTM is to the RDM. As we approach the diagonal line ($l=u$), the GT transform approaches a hard threshold, emphasizing the topology rather than the geometry. As we move diagonally from the bottom left to the upper right, the RGTMs go from emphasizing the local neighbor relationships among the stimuli to emphasizing the global structure of the representation.
}\label{fig3}
\end{figure}

\subsection{Nonlinear monotonic transforms of representational dissimilarities provide a family of geo-topological descriptors}

Topological RSA builds on the literature on topological methods (e.g., persistent homology \cite{zomorodian2005topology} and TDA mapping \cite{carlsson2009topology}). In order to suppress noise, we would like to find a lower threshold $l$ below which we consider stimuli as co-localized (i.e. the distance is $0$). In order to abstract from idiosyncrasies of individual brains and highlight the representational properties that are key to their computational function, we would like to find an optimal upper distance threshold $u$ above which we consider stimuli maximally distinct (i.e. we do not consider differences between larger distances meaningful). Two stimuli whose distance is larger than $u$ are disconnected in the graph capturing the topology.

Between the two thresholds we place a continuous transition so as to retain geometrical sensitivity in the range where it is meaningful (Fig. \ref{fig3}a). For simplicity, we propose a piecewise linear function as the monotonic distance transform. Note, however, that alternative monotonic transforms, such as the logistic function or a cosine transition, could be applied here. Given an original distance $d_{i,j}$ between the two neural signatures of two objects in the representational space, the piecewise linear geo-topological (GT) transform is defined as:

\begin{equation}
GT_{l,u}(d_{i,j}) = 
\begin{cases}
    0,& \text{if } d_{i,j} \leq l\\
    \frac{d_{i,j}-l}{u-l}, & \text{if } l < d_{i,j} < u\\
    1, & \text{if } u \leq d_{i,j} 
\end{cases}
\label{gt}
\end{equation}

By varying the lower bound $l$ and upper bound $u$, we obtain a family of GT transforms (Fig. \ref{fig3}b). Each member of this family transforms the original dissimilarity matrix (RDM) into a \textit{representational geo-topological matrix (RGTM)}, which provides a multivariate summary statistic with particular degrees of topological and geometrical sensitivity (Fig. \ref{fig3}c). The RGTM replaces the RDM in topological RSA. Note that the RDM is itself a member of the family, where $l=0$ and $u$ is set to the maximum (upper left corner in Fig. \ref{fig3}b, c, d). The RGTM, thus, generalizes the RDM.

When applying topological RSA to neural data to understand a brain representation, we can benefit from considering not only the RDM, but also other members of the RGTM family to gain an understanding of the geometrical and topological features of a brain representation. 
For the particular purpose of model selection, we aim to choose thresholds $l$ and $u$, so as to best recover the true (data-generating) model representation if it is among our models, or the best approximation among the models we are testing. 

\subsection{The geo-topological descriptor family captures both geometric and multi-scale topological information}

For interpretation of the family of GT transforms, we can consider the two diagonal dimensions of the triangular set spanned by $l$ and $u$ (Fig. \ref{fig3}d): From the upper left corner (RDM) toward the lower right, we go from geometrical to topological sensitivity, approaching the thresholding transforms ($l=u$) along the diagonal. From the lower left to the upper right, we go from local to global distance sensitivity.

For instance, the original RDM is in the top left corner ($l=0, u=1$), and around it, we have a region of RGTMs that is \textit{geometry-sensitive} where $l$ is small and $u$ is big. On the bottom left corner, where both $l$ and $u$ are small, we have \textit{local extractors}, which are RGTMs that are sensitive to whether or not two items are very close neighbors. On the upper right corner, where both $l$ and $u$ are big, we have \textit{global extractors}, which are RGTMs that are sensitive to whether or not two items are on opposite ends of the ensemble. If we go along the diagonal line from lower left to upper right, we have a belt zone of RGTMs which are close to binary (i.e. $l$ and $u$ are close to each other).

Formally, we require $l<u$, so the diagonal, where the transform is a simple thresholding function, is excluded from the family. Allowing $l=u$, so as to formally include simple thresholding functions, is possible but would complicate Eq. \ref{gt}. In either variant, the GT transform approaches a binary thresholding function for points approaching the diagonal. The thresholding functions along the diagonal relate our approach here to the mathematical filtration process used to reveal persistent homology in topological data analysis.

By exploring the choices of $l$ and $u$, we can identify the GT transforms that best enable us to match functionally corresponding cortical areas between different individuals. Similarly, we can generate data (with simulated measurement noise) from a layer in a deep neural network model, and determine which choices for $l$ and $u$ best enable us to identify the data-generating layer when using a range of layers from other instances of the DNN architecture (trained from different random seeds) as the models in analyses. 

One possibility is that the ideal setting is $l=0$, $u=max$, i.e., the original RDM, which characterizes the geometry. Other settings of $l$ and $u$ remove information about the geometry. Whether removing information by choosing a larger $l$ or a smaller $u$ helps or hurts depends on the relative extent to which it reduces signal and noise (nuisance variation) in the context of a particular data-analytical objective. If a topology-sensitive summary statistic reduced the variation caused by measurement noise and individual idiosyncrasies (i.e. nuisance variation) more than variation reflecting computational roles of different representations, then inferential comparisons of deep neural network models would benefit from topological RSA.

\subsection{Geodesic distances provide an alternative geo-topological descriptor}

In addition to RGTMs, we propose the use of geodesic distances in the representation. Let us first consider the theoretical notion of a geodesic and then the practical analyses it motivates. Theoretically, the ``representation'' of a population of possible stimuli can be defined as the set of response patterns the stimuli elicit. If the stimulus population is a continuous set, we may hypothesize that so is the set of corresponding neural response patterns in our brain region of interest. This set of neural response patterns is often referred to as the neural manifold. (Note, however, that the set of response patterns would need to be locally homeomorphic to a Euclidean space to conform to the definition of manifold. Whether this is the case for a particular neural population is an empirical question.) A geodesic distance between two stimulus representations is the length of a shortest path traversing the representational set from one to the other. Unless the straight line between the two points is a subset of the representation, we will traverse a longer, curved shortest path through the representational set. The length of the geodesic path then will be larger than the Euclidean distance.

In practice, we will have data for a finite sample from the population of stimuli. To estimate the geodesic distance on the manifold, we can measure geodesics in the discrete graph characterizing our representation. In a graph, a geodesic distance is the length of the shortest path between two nodes. We define a representational graph for each member of the RGTM family. For each node $i$, edges exist only to other nodes $j$ with dissimilarity $d_{i,j}<u$. The edge weights are defined by the transform $GT(l,u)$ and are interpreted as distances between nodes. Note that edge weights, thus, can be zero. Zero-distance edges can be motivated as a correction of small positive distance estimates resulting from off-manifold noise displacements of patterns whose locations on the manifold are not significantly distinct. In order to maintain direct comparability of the edge-weight matrices of different brain and model representations, we do not collapse nodes with zero distance, but maintain one node for each stimulus.

The geodesic distance between two nodes is defined as the length of the shortest path that leads from one node to the other, where the length of a path is the sum of the internode distances (i.e. the edge weights) along it. The geodesic distance is infinite if there is no path connecting the two nodes through the edges. The shortest paths for all pairs of nodes (each corresponding to a stimulus) can be found using Dijkstra's algorithm \cite{dijkstra2022note}. The result is a stimulus-by-stimulus matrix, which we refer to as the \textit{representational geodesic-distance matrix (RGDM)}. The RGDM can be used in place of the RGTM or the RDM for our visualization and model-comparative inference procedures. As an alternative to an RGTM-based graph, we could use a binary graph (edge weights $\in {0, 1}$) to compute the RGDM. For example, we could use the binary graph in which each node is connected to its $k$ nearest neighbors or the graph containing connections for the $k$ smallest representational dissimilarities in the RDM. Our analyses here, however, use RGDMs computed from graphs with continuous internode distances, based on members of the RGTM family as described above.

\subsection{Leave-one-out evaluation on human fMRI data and DNN models can quantitatively evaluate the region identification power of the summary statistics}

We would like to quantitatively evaluate the power of topological RSA in the context of model selection, where each model predicts a representational geometry.
We therefore consider cases, where the ground-truth model is known. This enables us to objectively evaluate the impact of different choices of $l$ and $u$ on the accuracy of model selection. If conventional RSA were optimal, then setting $l=0$ and $u=max$ would be the best settings. If other settings afforded equal or better accuracy for model selection, then topological signatures would deserve consideration in future studies applying RSA to adjudicate even among models that predict not just representational topologies, but full representational geometries.

\textbf{Evaluating topological RSA's brain-region-identification accuracy (fMRI).} The fMRI evaluation was performed on pre-existing data from a human fMRI experiment \cite{walther2015beyond,walther2016sudden}, in which 24 subjects were presented  with 62 colored images depicting faces, objects, and places. We use 8 regions of interests (ROIs) here: the primary visual cortex (V1), the secondary visual cortex (V2), the extrastriate visual cortex (V3), the lateral occipital complex (LOC), the occipital face area (OFA), the fusiform face area (FFA), the parahippocampal place area (PPA), and the anterior temporal lobe (aIT).  

We investigate the brain-region-identification accuracy (RIA), where each brain region is considered a model. The region labels provide the ground truth: For data from each brain region in a held-out subject, we would like to identify which region the data came from on the basis of the data for all the regions from the other subjects. We therefore perform leave-one-subject-out (LOSO) RIA evaluation. First, we randomly sample 10 sets of $l$'s and $u$'s in each of the five interpretable RGTM zones: topology-sensitive (TS), geometry-sensitive (GS), local extractor (LE), global extractor (GE), and intermediate (I). The 10 samples of $l$ and $u$ for each region define 10 different GT transforms and provide an estimate of the RIA, averaged across regions and subjects, that reflects the performance at region identification of different zones of the family of RGTMs.

For each region in a held-out subject, we assign the region label of the average RGTM from the other subjects that is closest (in terms of Euclidean distance) to the RGTM being identified. Each subject is held out once in a full crossvalidation cycle and the RIA is the average identification accuracy.

In order to inferentially compare the different zones of the RGTM family, we perform frequentist comparisons. We would like to consider a difference in performance as significant if we expect it to generalize to experiments performed with different samples of subjects and stimuli drawn at random from the same populations of subjects and stimuli. We therefore perform a 2-factor bootstrap procedure, resampling both subjects and stimuli simultaneously \cite{nili2014toolbox, schutt2023statistical}. The standard deviation of the RIA estimates across 1,000 bootstrap samples of subjects and stimuli serves as our estimate of the standard error of the RIA estimate. Two-sided t-tests are then applied to assess the significance of differences between RIA estimates for different choices of $l$ and $u$. The degrees of freedom in this approach correspond to the smaller factor, which is the number of subjects (24) in this case.

\textbf{Evaluating topological RSA's layer identification accuracy (DNNs).} The DNN evaluation was performed on a convolutional neural network architecture \cite{lecun1995convolutional} called the All Convolutional Neural Net \cite{springenberg2014striving}. We investigate the layer identification accuracy (LIA), where each layer is considered a candidate brain-computational model. We perform leave-one-instance-out (LOIO) LIA evaluation. We trained 10 model instances, starting from 10 different random seeds, of the All-CNN-C network architecture \cite{springenberg2014striving}, a 9-layer fully convolutional network
that exhibits state-of-the-art performance on a well-known small object-classification benchmark
task (CIFAR-10 \cite{krizhevsky2009learning}), .

We used the same numbers of feature maps (96, 96, 96, 192, 192, 192, 192, 192, 10) and kernel dimensions (3, 3, 3, 3, 3, 3, 3, 1, 1) as in the original paper. The training of All-CNN-C network instances involved 350 epochs using the ADAM optimizer with a momentum value of 0.9 and a batch size of 128. A preliminary learning rate of 0.01 was employed, along with an L2 regularization coefficient of $10^{-5}$ and gradient norm-clipping value of 500. Following \cite{mehrer2020individual}, we trained the DNNs on the complete CIFAR-10 image dataset (both training and test sets), which comprises 10 distinct object categories, each represented by 5000 training and 1000 test images, implemented with TensorFlow (version 1.3.0) and Python 3.5.4.

Like different individual subjects, these instances differ in their detailed connectivity, but perform the recognition task at similar levels of accuracy \cite{mehrer2020individual}. In addition to evaluating the LIA across instances for different choices of $l$ and $u$, we study the effect on the LIA of injecting Gaussian noise of a variety of variances $\sigma^2$ into the dissimilarity estimates.
% and the Bernoulli noise $\epsilon$ (applied during training as dropout rates).

The DNN-simulation-based evaluations follow the same procedures as the human-fMRI-based evaluations: The neural networks were presented with the same set of 62 object images to define the representational geometries. We perform LOIO LIA evaluation. Each layer in a held-out instance is identified as the layer whose average RGTM across the other instances is closest (in terms of Euclidean distance) to the RGTM being identified. Each instance is held out once in a full crossvalidation cycle and the LIA is the average identification accuracy. The inference, likewise, employs the same 2-factor (instance and stimulus) bootstrap method.

\subsection{Model-comparative statistical inference for tRSA}

Topological RSA (tRSA) can use the well-developed inferential techniques of RSA for comparison between representational models \cite{kriegeskorte2008representational, diedrichsen2020comparing, schutt2023statistical}. The nonparametric inference methods of RSA3 \cite{schutt2023statistical} can simply use the geo-topological statistics (RGTMs and RGDMs) in place of RDMs. As in conventional RSA, the representational similarity of brain regions and model layers can be quantified using various comparators, such as cosine similarity or correlation, but applied to geo-topological statistics rather than RDMs.

Here we used Euclidean distance as the comparator for the geo-topological summary statistics. The $l$ and $u$ are defined as quantiles (expressed as percentiles or ranks) relative to the set of dissimilarities in a given RDM. Defining $l$ and $u$ as quantiles enables matching choices for model and brain representations whose dissimilarities may have different magnitudes and may lack a common unit that would render them commensurable. In addition to defining $l$ and $u$ as quantiles, we use the ranks within each RDM to define the dissimilarities entering the GT transform. This has the benefit that the resulting RGTMs have identical distributions of values. In this scenario, the squared Euclidean distance as a comparator of representational summary statistics is proportional to the Pearson correlation distance and to the cosine distance. These comparators thus would have yielded identical model-selection results, rendering the analyses relevant to the most common choices of RDM comparator in RSA.

Another popular choice for the RDM comparator is a rank correlation coefficient, such as Kendall's $\tau_a$ \cite{nili2014toolbox} or the more computationally efficient Spearman-type coefficient $\rho_a$ \cite{schutt2023statistical}. In the present study, the RDMs are rank-transformed before computing the RGTMs. Our comparators therefore benefit from the robustness afforded by the rank transform, obviating the need for another rank-transform at the level of the RGTMs, as would happen if we chose a rank correlation coefficient as the comparator. Using a rank correlation coefficient is closely related (but not mathematically equivalent) to the present analyses.

\begin{figure}[tb]
% \vspace{-0.8cm}
\centering
    \includegraphics[width=0.9\linewidth]{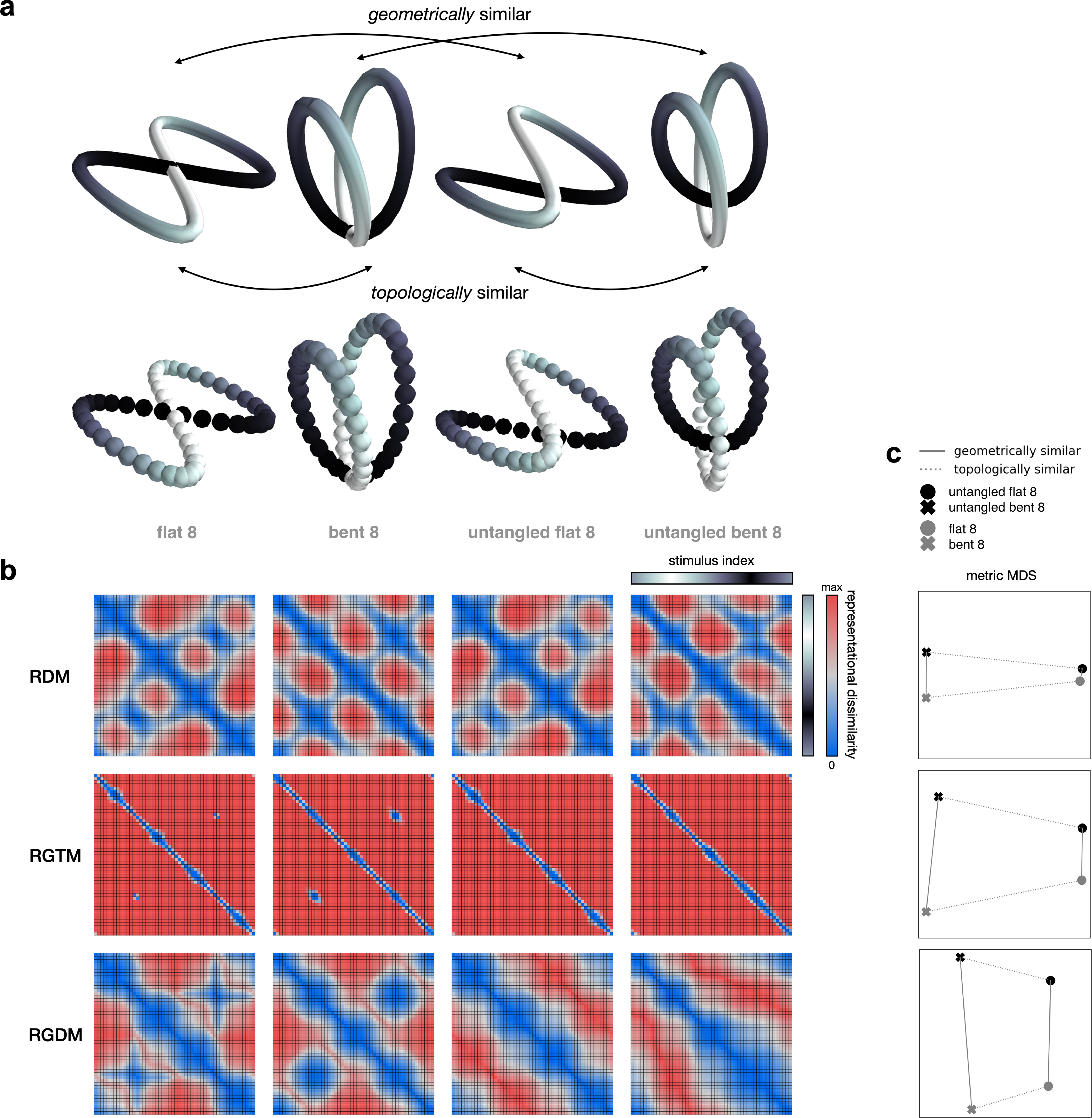}
\caption{\footnotesize \textbf{The geometric and topological similarities between hypothetical neural representations (proof of concept).} \textbf{(a)} We consider four hypothetical representations of 40 stimuli (balls in lower row of a). The response patterns are sampled from idealized continuous sets of neural response patterns (top row in a). Two of these continuous sets are manifolds (the untangled shapes) and the other two are not (the tangled shapes, where the set self-intersects, forming a neighborhood that is not homeomorphic to a Euclidean space). From left to right, we label the four representations the flat 8, the bent 8, the untangled flat 8, and the untangled bent 8. The flat 8 and the untangled flat 8 are geometrically similar, while being topologically dissimilar (with the former self-intersecting). Their bent versions, as well, are geometrically similar and topologically dissimilar. The flat 8 and the bent 8, on the other hand, are topologically similar (with the self-intersection creating two holes) and geometrically dissimilar (because the twisting substantially changes the metric geometry). Their untangled versions, as well, are topologically similar and geometrically dissimilar. \textbf{(b)} The RDMs (Euclidean distance, top row) do a good job characterizing the geometric relationships but are insensitive to the topological relationships. The RGTMs (middle row) are more sensitive to topology. To achieve local topological sensitivity, we chose  $l=0$ and $u=0.075$, revealing the self-intersection in the two leftmost representations. The RGDMs (bottom row) are exquisitely sensitive to the topological relationships, while de-emphasizing geometrical relationships between the representations. Note the ``eyes'' in the two leftmost RGTMs, representing the self-intersections. Each RGDM captures the lengths of the shortest paths in the graph of the RGTM shown above it. The RGDMs more prominently reflect the topology as the ``shortcut'' paths enabled by the self-intersection affects shortest-path lengths for a broad swath of stimuli. \textbf{(c)} Multi-dimensional scaling (MDS) on the four representations shows the pairwise similarities among the four matrices of each row, confirming the increasing sensitivity to topological differences as we go from RDM to RGTM and on to RGDM.   
}\label{fig4}
\end{figure}

\begin{figure}[tb]
% \vspace{-0.8cm}
\centering
    \includegraphics[width=\linewidth]{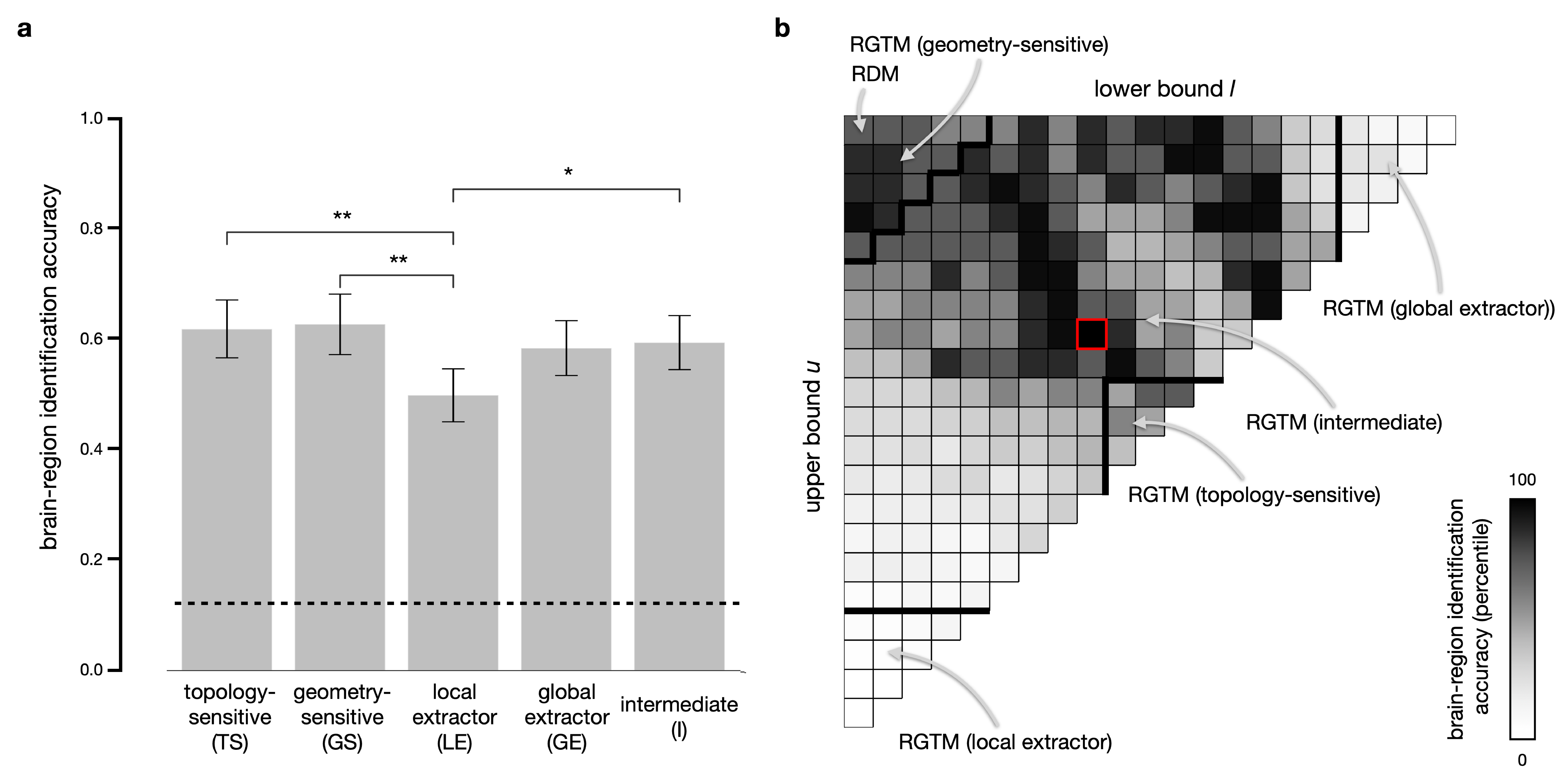}
\caption{\footnotesize \textbf{Brain-region identification accuracy across the family of geo-topological descriptors.} Our analysis of human brain regions used fMRI data from 24 subjects, collected from 8 regions of interest, including the primary and secondary visual cortices, the lateral occipital complex, the occipital face area the fusiform face area, the parahippocampal place area, and the anterior temporal lobe. (\textbf{a}) Region-identification accuracy (RIA) is evaluated using leave-one-subject-out crossvalidation, where a classifier (on which we compute RIA) is trained on all available data except for one subject and then tested on that left-out subject. This process is repeated for each subject, with the final performance measure being the average across all iterations. We used bootstrapping to obtain an unbiased estimate of the standard error as the error bound for region identification accuracy and and used crossvalidation to prevent overfitting.  We randomly sampled 10 sets of lower bounds $l$'s and upper bounds $u$'s in each of the five interpretable RGTM zones as defined in Fig. \ref{fig3} and applied a paired \textit{t}-test to compare the RIA between different RGTM zones as defined in panel b) and Fig. \ref{fig3} (**** p < 1e-4, *** p < 1e-3, ** p < 1e-2, * p < 0.05). \textbf{(b}) RIA percentile (gray colorscale) as a function of the combination of upper and lower bounds, $l$ and $u$, defining the RGTM (layout as in Fig. \ref{fig3}) with the best-performing RGTM marked in red.}\label{fig5}
\end{figure}

\begin{figure}[tb]
% \vspace{-0.8cm}
\centering
    % \vspace{-8mm}
    \includegraphics[width=0.95\linewidth]{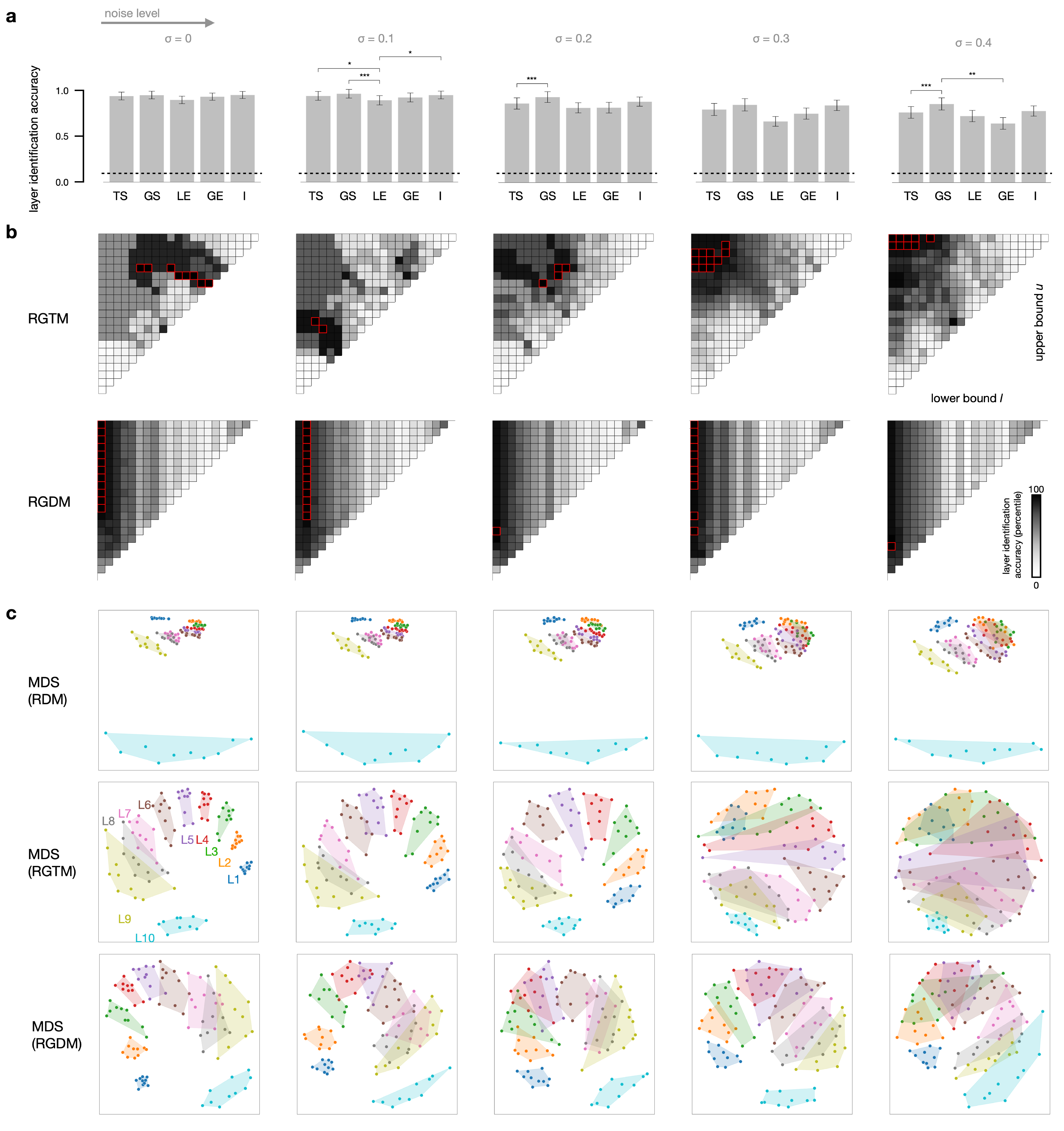}
    % \vspace{-6mm}
\caption{\footnotesize \textbf{Neural-network-layer identification accuracy across the family of geo-topological descriptors.} The performance of different RGTMs at neural-network layer identification with gradually increasing Gaussian noise $\sigma$ on the dissimilarity estimates (columns). Results are for All Convolutional Neural Nets, a simplified model architecture with many convolutional layers (with 10 separate layers of interest in total). Layer identification accuracy (LIA) was estimated in leave-one-instance-out cross-validation, where a classifier is trained on all available model instances except for one  and then tested on that left-out model instance. This process is repeated for each instance, with the final performance measure being the average across all iterations. To enable crossvalidated evaluation similar to the brain-activity data in the previous figure, for each model, we trained 10 instances (analog of individual subjects) from different random seeds. Cross-validation was performed within a bootstrapping process to estimate LIA and its estimation error. We randomly sampled 10 sets of $l$'s and $u$'s in each of the five interpretable RGTM zones as defined in Fig. \ref{fig3} and applied a paired \textit{t}-test to compare the LIAs afforded by different RGTM zones. \textbf{(a)} Bar graphs showing the LIA for each RGTM zone and noise level with statistical comparisons. LIA is similar for topology-sensitive RGTMs and geometry-sensitive RGTMs. \textbf{(b)} The LIA of RGTMs and corresponding RGDMs across the $(l,u)$ threshold pairs are represented in a percentile heatmaps, with the maximum performance marked in red. \textbf{(c)} The MDS plots of the RDMs (top row) show geometrical similarities among layer representations. Each dot is the RDM for one layer in one model instance. For each layer, a convex hull is drawn to group its representations across all instances. MDS plots reflecting geo-topological similarities are shown for the best-performing RGTMs and RGDMs (red in b), revealing the good discrimination of layers these geo-topological summary statistics provide (middle and lower rows, respectively). 
}\label{fig6}
\end{figure}

\section{Results}

\subsection{Proof of concept: geo-topological descriptors can discern topological and not just geometric  distinctions}

As a proof of concept, consider the geometric and topological similarities among the four hypothetical representations in Fig. \ref{fig4}. We call the four representations the ``flat 8'', the ``bent 8'', the ``untangled flat 8'' and the ``untangled bent 8''. Imagine an experiment in which the idealized continuous representational set is sampled using of 40 stimuli (balls in Fig. \ref{fig4}a). The flat 8 and the bent 8 share the self-intersection, which creates two holes, rendering the shapes topologically similar, although the bending of the latter greatly changes the geometry. Similarly, the untangled flat 8 and the untangled bent 8 are topologically similar (in that neither has a self-intersection and both have one hole) and geometrically dissimilar because of the bending.
By contrast,  the flat 8 and the untangled flat 8 are geometrically similar (both flat and similar in their RDM), and topologically distinct (one versus two holes). Likewise, the bent 8 and untangled bent 8 are geometrically similar (both bent) and topologically distinct (double arrows in Fig. \ref{fig4}a).

The different summary statistics reflect the topology and geometry of these hypothetical neural representations to different degrees (Fig. \ref{fig4}b). As expected, the RDM reflects the geometric distinctions but is not very sensitive to the topological distinctions, which are implemented here through minimal metric displacements that determine whether or not there is a self-intersection. The multi-dimensional scaling (MDS) arrangement of the four RDMs (Fig. \ref{fig4}c) shows the dominance of the flat versus bent distinction when characterizing the representations with RDMs.

The representational geo-topological matrices (RGTMs) were obtained using small values for both $l$ and $u$, yielding almost binary matrices with local topological sensitivity, which reveal whether or not there is a self-intersection. The MDS shows that the RGTM here balances sensitivity to the topology and the geometry of the representation. Note the prominent blue ``eyes'', which reflect the self-intersection where the contour of the 8 crosses itself. The blue ``eyes'' are present for the two-hole representations with self-intersection (left two representations in Fig. \ref{fig4}b), but absent for the untangled one-hole representations. 

Finally, the representational geodesic matrices (RGDMs) are even more sensitive to the topological distinctions. The RGDMs were defined on the basis of the graphs of the RGTMs in the row above. The shortest-path lengths reflect the existence of ``shortcut'' routes through the self-intersection. For many pairs of stimuli, the shortcut (when available) provides a shorter path than going the straight route around the 8. The small geometrical distortion causing the self-intersection is therefore reflected across a large portion of the RGDM. The MDS plots show that the RGTMs and RGDMs can effectively characterize both the geometric and topological similarities, whereas the RDMs mostly reflect the geometry of the representation.

\subsection{Geo-topological descriptors reveal what aspects of geometry and topology enable accurate identification of brain regions across subjects (fMRI)}

Using human fMRI data, we quantitatively evaluated the performance of the RGTM summary statistics (including the RDM as a special case) at revealing the correspondences among ventral-stream cortical regions across subjects. A summary statistic will succeed in this evaluation, if it is robust to noise and interindividual variability while maintaining sensitivity to differences between different cortical areas.

We grouped the members of the RGTM family into interpretable zones in Fig. \ref{fig3}d. Results for each zone are shown in Fig. \ref{fig5}a). We plotted the region identification accuracy (RIA) as a heatmap across choices for $l$ and $u$ (Fig. \ref{fig5}b). The highest RIA was achieved for $l=0.40$ and $u=0.65$ (red square in Fig. \ref{fig5}b). However, other settings yielded similar RIA. Results suggest that $l$ and $u$ should not be both low or both high. The RDM is an effective summary statistic. However, it contains information not needed for region identification. This result suggests that compressing less informative distance variation is possible without a reduction in model-selection accuracy. In this data set, we did not find that the benefits of noise reduction significantly outweighed the loss of information.

We inferentially compared RIA between different zones within the RGTM family using simultaneous bootstrapping of both subjects and stimuli (2-factor bootstrap, \cite{schutt2023statistical}) for frequentist pairwise comparisons. The 2-factor approach serves to test for differences expected to generalize across random samples of subjects and stimuli drawn from the same populations. The summary statistics in the topology-sensitive (TS) and geometry-sensitive (GS) zones did not perform significantly differently ($p = 0.434$). The topology-sensitive (TS), geometry-sensitive (GS), and intermediate RGTMs all performed better than those in local extractor (LE) zone ($p = 4.776e-03$, $p = 4.972e-03$,  and $p = 4.454e-02$, respectively). We found no other significant differences. Overall, geometry-sensitive and topology-sensitive summary statistics performed similarly, but the latter suffer when the upper bound is in the lower third of the distribution.

These findings suggest that a substantial portion of the smallest and largest dissimilarities can be compressed without reducing the RIA. Using the RGTM to focus on the intermediate range of dissimilarity variation (perhaps the middle third of the distribution of dissimilarities) while compressing the lower and upper third of the dissimilarities yields optimal model selection in this data set.

\subsection{Geo-topological descriptors reveal what aspects of geometry and topology enable accurate identification of DNN layers across instances}

Deep neural networks (DNNs) have emerged as important models of vision and brain-information processing in recent years \cite{yamins2016using,cadieu2014deep,long2018mid,kriegeskorte2015deep,dwivedi2021unveiling}. DNNs can learn non-linear representational transformations through feedforward and recurrent processing, enabling them to capture the computations underlying task performance. When a neural network architecture is trained repeatedly, starting from different random connection weights, the resulting trained model instances have distinct parameters and distinct detailed unit tuning, despite performing the same task roughly equally well. Just like individual humans differ, thus, so do instances of DNNs trained from different random seeds \cite{mehrer2020individual}. 

Analogously to the brain-region identification analysis in the previous section, we investigated the usefulness of RGTMs in identifying which layer of a neural network model a given RGTM was computed from, given RGTMs for all layers in other instances of the architecture. Like the human fMRI analysis, this analysis informs us about the best choice of representational summary statistic when performing DNN model comparisons using human data. We are able to objectively evaluate the summary statistics, because we know the ground-truth data-generating model. The DNN simulation analyses are complementary to the human fMRI analyses: On the one hand, they afford complete knowledge of the computational mechanisms (an advantage). On the other hand, they are less realistic with respect to the biology and structure of the noise in the data used as a basis for model comparison (a disadvantage).

Our layer-identification analysis reveals to what extent different RGTMs (including geometry-sensitive and topology-sensitive members of the family of GT transforms) abstract from nuisance variation across the same layer in different model instances, while capturing computationally meaningful variation between different layers (which are thought to play distinct computational roles).

As for the human brain regions, we quantitatively evaluated the performance of the summary statistics from the RGTM family (including the RDM). Results are shown in Fig. \ref{fig6}, grouped by noise level (Fig. \ref{fig3}d). We present results for two types of noise: Gaussian noise $\sigma$ (added to the dissimilarity estimates, Fig. \ref{fig6}) and Bernoulli noise $\epsilon$ (applied during training as dropout rates, Fig. S\ref{fig7} in the Supplementary Materials).

Statistical comparisons of layer identification accuracy (LIA) for GT-transform zones showed that geometry-sensitive statistics outperformed topology-sensitive statistics at higher noise levels ($\sigma \geq 0.2$). The heatmaps showing LIA as a function of $l$ and $u$ in RGTMs (Fig. \ref{fig6}b) likewise show that as the noise level increases, the optimal statistics (red frames) go from more topology-sensitive to more geometry-sensitive. The LIA heatmaps based on the RGDMs show a prominent fall-off of accuracy with $l$ (horizontal axis, as in Fig. \ref{fig3}) and little dependence on $u$: As $l$ grows, more and more pairs of neighboring points collapse to $0$ distance in the RGTM, and thus more and more shortest-path lengths between points connected by paths of $0$-distance edges collapse to $0$ in the RGDM. The weak dependence of LIA on $u$ reflects the fact that more separated points can be connected either through a smaller number of larger steps (when $u$ is large) or a larger number of smaller steps (when $u$ is small), rendering the RGDM less sensitive than the RGTM to $u$. Overall, in RGTMs, there is no significant LIA difference between the geometry-sensitive and the topology-sensitive zone when $\sigma < 0.2$. Selecting a summary statistic that emphasizes only local neighborhood (local extractor) can be significantly worse than selecting the geometry-sensitive or intermediate statistics, consistent with what we observed for the human fMRI data. This suggests that, in a low-noise regime, some of the information in the RDM is redundant and the extremes of the distance distribution can be compressed without compromising our power to distinguish representations of different computational modules (layers here). In the high-noise regime, having the full RDM information can better characterize the subtle distinctions among computational modules.

To further interpret the distributions of representations associated with different layers and instances, we performed multi-dimensional scaling (MDS) on the RDMs, RGTMs, and RGDMs (using the optimal threshold pairs $(l,u)$ marked in red in the corresponding heatmaps in Fig. \ref{fig6}b). We used the Procrustes alignment \cite{procrustes} to obtain a consistent orientation and reflection of the MDS plots (Fig. \ref{fig6}c). We observe that the RDMs cluster the representations of several layers together, whereas the RGTMs and RGDMs manage to separate out layers in a smoother progression from the early layers to the later layers. This suggests that the geo-topological and geodesic transforms can help emphasize topological invariants that consistently characterize layer representations across variation among model instances.

\section{Discussion}

Theories about neural codes and computations can predict how the neural manifold hangs together (i.e. the topology) without predicting a specific representational geometry. Such theories are consistent with an infinite number of distinct geometries, and so do not predict a specific RDM. Topological RSA enables researchers to evaluate and statistically compare such theories as expressed by a representational graph and associated RGTM. Our results do not support the conclusion that topological descriptors should replace geometrical descriptors in RSA in general. Studies that aim to adjudicate among brain-computational models that predict representational geometries can continue to use geometrical descriptors. Topology may not be needed as a tool to suppress nuisance variation in RSA analyses of representational geometries. However, the topology of neural representations is interesting in its own right.

For theories and brain-computational models that do predict geometries, evaluation on the basis of their topological predictions provides an alternative and complementary perspective. If our models' RDM predictions do not reach the noise ceiling, the model that best predicts the geometry may not be the model that best predicts the topology. We can simply apply a geo-topological transform to the brain RDM before performing model-comparative RSA inference with a rank-based comparator. The choice of parameters $l$ and $u$ (which can be defined as percentiles within the dissimilarities of each RDM) enables us to calibrate the relative sensitivity to the topological and geometrical properties of the representation. This approach enables comparisons of both topology-predicting and geometry-predicting models within the same tRSA framework. The fact that region identification and layer identification do not suffer when we drastically reduce the information in the RDM suggests that the bulk of the information distinguishing the representations in brain regions and NN layers is captured by topological descriptors.

\subsection{Characterizing topology and geometry provides a comprehensive view of neural representations} 
%Topology-sensitive representational summary statistics offer a unique perspective on neural representations by capturing the underlying topological structure of a set of neural population response patterns. These statistics can reveal relationships in the data that are not apparent when using traditional geometric descriptors alone. By considering the topological aspects of neural representations, we may be able to focus on the invariants of particular cognitive algorithms, abstracting from noise and individual idiosyncrasies. 

Considering topology-sensitive representational summary statistics alongside geometrical ones can provide a more comprehensive and nuanced understanding of the data. We introduced geo-topological descriptors as a new class of summary statistics for RSA. These descriptors emphasize the topology by compressing variation among the smallest representational distances and among the largest representational distances. Variation among small representational distances is sensitive to noise, and variation among very large distances may reflect individual idiosyncrasies more than computationally meaningful information. While tRSA did not outperform conventional RSA at region or layer identification, it matched the model-selection performance of RSA while substantially reducing the information, thus revealing what range of variation among representational distances captures the discriminable computational signatures of different representations. Topological RSA provides a robust way to identify  neural representations in spite of noise and interindividual variation. We emphasize the synergistic potential of combining the topological and geometrical perspectives. 

\subsection{Testing topological representational hypotheses requires tRSA} 

From a theoretical perspective, the choice of using topological or geometrical summary statistics in RSA depends on the hypotheses about the neural representations that are to be tested. If the hypothesis is that a neural representation conforms to a particular topology, then topological RSA provides a straightforward method to test the hypothesis. Note that a hypothesis about the representational topology cannot be straightforwardly tested with conventional RSA because the hypothesis corresponds to a complex set of RDMs any of which conforms to the hypothesized topology. This is a major motivation for tRSA independent of the question of whether it can help reduce nuisance variation in adjudicating among models that predict geometries.

%One direction for future work is to clarify how topological descriptors relate to known neurophysiological properties and computational principles. 
Do theories about neural mechanisms imply specific predictions about the representational topology? Investigating the relationship between brain-computational theories and the topology of neural population representations is an important direction for future computational work. 
%Moreover, geo-topological analyses of computational models can generate testable biological hypotheses for future experimental studies.
Here we focused on visual representations in both the model-based simulations and the fMRI dataset, where the representational space has many dimensions and we do not have simple topological hypotheses about the structure of the representations. In the absence of simple topological hypotheses, a data-driven approach can reveal to what extent the representational topologies are consistent between individuals and distinct across stages of processing or across stages of learning or development. The relationships among brain-representational topologies can be visualized using MDS of the RGTMs or RGDMs as we show for DNN layers in Fig. \ref{fig6}c. Such analyses have been performed previously for RDMs (e.g. \cite{kriegeskorte2008representational, visconti2017neural} and require independent data for each region to prevent correlated noise fluctuations from confounding the RDM estimates as shown in \cite{henriksson2015visual}. Future studies could use tRSA to test strong predictions about the topology of lower-dimensional stimulus sets sampling, e.g., orientations and spatial frequencies of gratings, or directions and velocities of visual motion, which are known to be represented at different levels of the primate visual hierarchy. Another example where tRSA could support a strongly theory-driven approach seeking to adjudicate among alternative topological hypotheses is the head and travelling direction system in mice and fruit flies \cite{chaudhuri2019intrinsic,langdon2023unifying,kim2019generation,lyu2022building}.

\subsection{Testing geometrical representational hypotheses may benefit from tRSA}

If the models to be tested predict specific representational geometries, conventional RSA can be used to adjudicate among them. Topological descriptors are functions of geometrical ones and reduce the information, so may appear undesirable if geometrical hypotheses are to be tested. Consider the example of a set of visual gratings covering the entire cycle of orientations with equal spacing. For a wide variety of hypothetical representations, we expect the topology to be that of a circle, whereas the geometry may not be that of a circle, for example, if the neural code is anisotropic, more precisely representing cardinal orientations (which are more frequent in natural images) \cite{kriegeskorte2021neural,appelle1972perception,girshick2011cardinal}. The topology, in that case, fails to capture an important feature of the neural code. In other scenarios, however, a topological descriptor may capture the essential features of the neural code. If our topological descriptor reduces nuisance variation in the data more than the representational signatures that distinguish different models, it can support improved model-adjudication accuracy. For the visual representations investigated here in models (deep neural networks) and humans (fMRI), we did not find evidence for such an advantage. Future studies will reveal if a focus on the topology reduces the noise more than the signal in in other circumstances. 
 
 However, even if our models make predictions about representational geometries (as do DNNs) and tRSA does not improve model-adjudication power, it can still provide complementary information by revealing to what extent the topology implicit to the geometrical predictions enables model adjudication and which model best predicts the representational topology. If the best model fully explains the geometry, it will also correctly predict the topology. However, the relative performance of models that do not fully explain the geometry can change when we evaluate models by their topological predictions. Our analyses here showed that tRSA, using reduced topological summaries, can match the performance of geometrical RSA at model adjudication. This suggests that topological descriptors that compress the variation among small and among large representational distances capture the information essential to the computational signature of different representations.

\begin{comment}
\subsection{Generalizability and robustness of geo-topological descriptors worth further research}

While the results of our study demonstrate that tRSA maintains a robust identification accuracy for brain regions or computational layers of interest across different noise levels, it is crucial to consider the generalizability of these findings to other datasets and experimental conditions. Are the observed benefits of the geo-topological descriptors consistent across different datasets, task paradigms, and neuroimaging modalities? Replication studies using diverse datasets can help assess the generalizability of the proposed method. Additionally, it would be valuable to investigate the robustness of the geo-topological descriptors to variations in data quality, such as noise levels or different preprocessing techniques. Evaluating the method's performance across different experimental conditions and data characteristics will provide valuable insights into its applicability and reliability.
\end{comment}

\subsection{Practical considerations for implementing tRSA}

Topological RSA model comparison can rely on the RSA3 statistical inference framework \cite{schutt2023statistical}, which has been implemented in Python in the open-source \href{https://github.com/rsagroup/rsatoolbox}{RSA Toolbox}). The geo-topological descriptors introduced here and the RSA3 inference methods enable analyses based on a wide range of brain-activity data, including invasive neural recordings, fMRI, EEG, and MEG. A topological representational hypothesis can be expressed directly in a representational graph and associated RGTM. Alternatively, topological hypotheses can be derived from representational models that predict geometries, i.e., any descriptors of the experimental conditions. If the conditions correspond to visual stimuli, for example, representational models can be derived from properties of the images or of the objects depicted and their categories \cite{kriegeskorte2008matching}. In particular, topological hypotheses can be derived from brain-computational models, such as neural network models that implement candidate hypotheses about the computations performed by the brain and predict a geometry for each stage of representation \cite{kriegeskorte2015deep, yamins2016using, kriegeskorte2018cognitive}.

If the theories to be compared predict representational topologies (RGTMs), then we can avoid having to choose $l$ and $u$. We can compare the RGTMs predicted by the theories to the brain RDM directly using the $\rho_a$ rank correlation coefficient \cite{schutt2023statistical}. If the theories to be compared predict geometries (RDMs), but we intend to compare models in terms of their predictions of the topology of the brain representation, then we can convert the brain RDM to a brain RGTM by selecting $l$ and $u$ to choose the desired calibration of the summary statistic to the geometry and the topology of the brain representation. The brain RGTM can then be compared to the model RDMs using the $\rho_a$ rank correlation coefficient and model-comparative inference can proceed on this basis. Finally, if the goal is to assess whether the geometrical predictions of a model (the model RDM) significantly exceed its topological predictions (model RGTM, based on a choice of $l$ and $u$), we can use the $\rho_a$ rank correlations with the brain RDM to statistically compare the model's topological and geometrical predictions.

The values for $l$ and $u$ can be set by a priori considerations, e.g., on the basis of the level of graph connectivity predicted by the theories to be evaluated, or on the basis of results of earlier studies that use a different data set (e.g., this study). If different settings of $l$ and $u$ are to be explored in the analyses, it is important to avoid biases to the analyses that can result from selecting these two parameters \cite{kriegeskorte2009circular, hosseini2020tried}. One approach is to choose a small number of settings for $l$ and $u$ and add each resulting RGTM as a separate model in the model-comparative inference, where standard adjustments for multiple testing can then be used (controlling the family-wise error rate or the false-discovery rate as implemented in the \href{https://github.com/rsagroup/rsatoolbox}{RSA Toolbox}). Alternatively, an independent analysis, for layer- or brain-region identification accuracy, could inform the choice of $l$ and $u$. Importantly, whatever exploration or search procedures were employed in selecting $l$ and $u$ must be fully reported.

It is important to note that both topological and geometrical descriptors depend on the experimental conditions (e.g. stimuli) for which activity patterns are included in the analysis. However, topological descriptors can be even more sensitive to the set of conditions included. For example, the geodesic distance between the representations of two conditions depends on the other conditions included (which determine the shortest path), whereas the Mahalanobis distance does not.

In conclusion, tRSA is essential for testing topological representational hypotheses. Even when testing models that predict representational geometries, tRSA can reveal to what extent the models capture, in particular, the representational topology observed in a neural population. The combination of topological and geometrical descriptors offers a promising comprehensive approach for analyzing neural representations.

\clearpage

\bibliographystyle{unsrt}
\bibliography{main}  

\clearpage
\appendix

\section{Supplementary Materials for ``The Topology and Geometry of Neural Representations''}

\begin{figure}[h!
]
% \vspace{-0.8cm}
\centering
    \includegraphics[width=\linewidth]{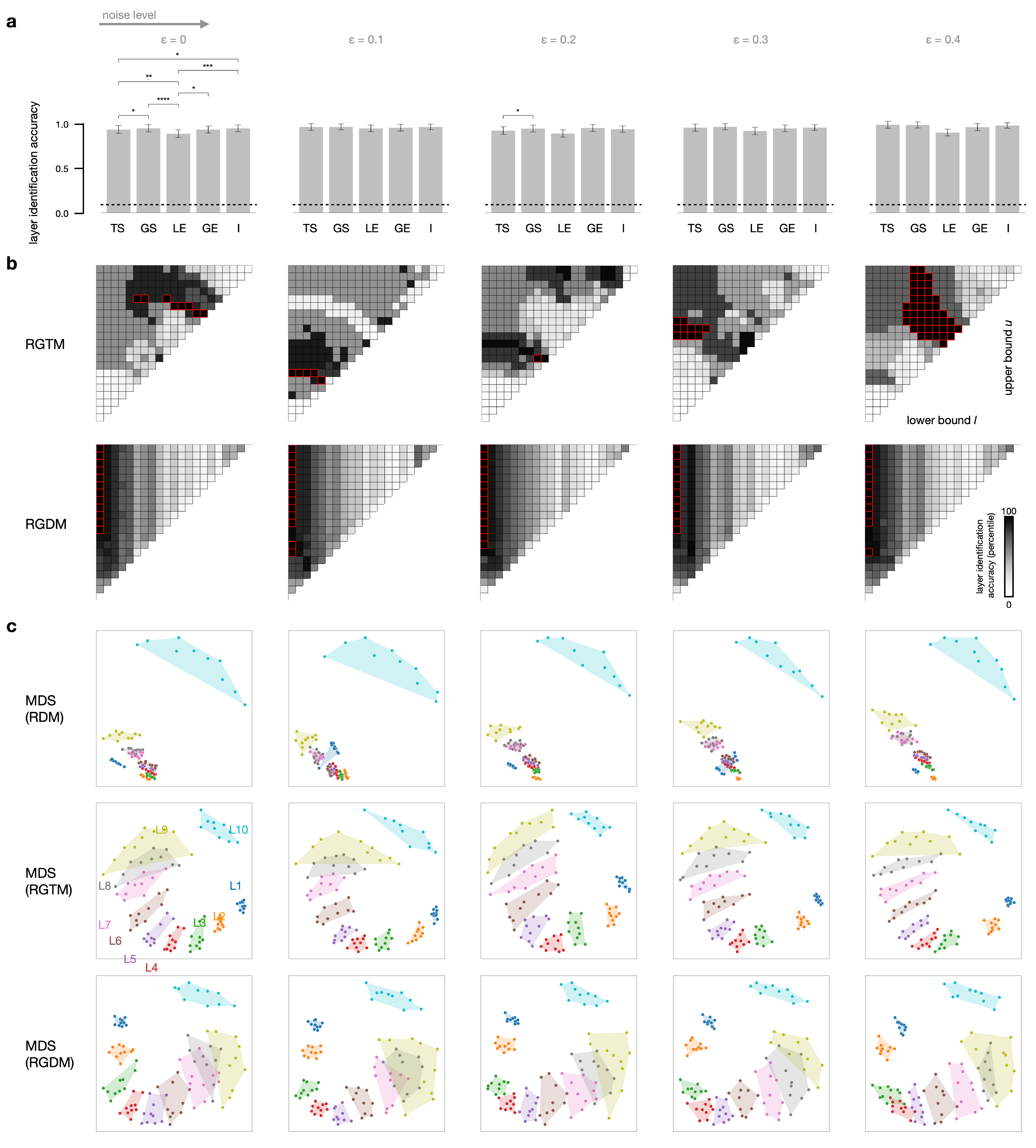}
\caption{\footnotesize \textbf{The performance of different RGTMs and their corresponding RGDMs of deep neural networks with gradually increasing Bernoulli noise $\epsilon$ during training.} Similar to Figure \ref{fig6}, the performance is evaluated by the layer identification accuracy (LIA) among the neural network layers in a leave-one-out cross-validation process, on the dataset of the same neural net architecture as Figure \ref{fig6}, but with gradually increasing Bernoulli noise $\epsilon$ instead of Gaussian noise.  
}\label{fig7}
\end{figure}

\end{document}